# The criterion on the Propulsive Balance of Oblique Detonation Engine

Yunfeng Liu[1] *, Xin Han [2], Zijian Zhang[3]

[1] Institute of Mechanics, Chinese Academy of Sciences, Beijing 100190, China

[2] School of Aerospace Engineering, Xiamen University, Xiamen 361102, China

[3] Department of Aeronautical and Aviation Engineering, The Hong Kong Polytechnic University, Kowloon 999077, Hong Kong, China

( Corresponding author: Yunfeng Liu, liuyunfeng@imech.ac.cn, )

**Abstract :** The oblique detonation engine (ODE) has established a clear superiority for hypersonic flight because of its high thermal efficiency and compact structure. It has become the research hot spot all over the world in recent years. The aim of this study is to derive a criterion on the propulsive balance of ODE, from which we can find the key parameters governing the propulsive performance explicitly. A physical model of ODE is put forth, which consists of the inlet, the constant cross-section combustor and the divergent nozzle. The mathematical equations to calculate the thrust generated by the divergent nozzle and the pressure drag produced by the inlet are deduced. The net thrust of ODE is then obtained. The criterion shows clearly that the static temperature at the engine inlet exit is a very important parameter. The lower the inlet exit temperature is, the higher the specific impulse will be. The specific impulse of ODE with stoichiometric $H_2$/air mixture and hydrocarbon/air mixture are calculated by using these equations. The results show that ODE can obtain positive net thrust from Ma8 to Ma15.

**Key words:** Oblique detonation engine, ODE, Propulsive performance, Specific impulse

# 1 Introduction

Oblique detonation engine (ODE) is one of the most promising air-breathing engines for hypersonic flight because of high thermal efficiency and compact structure. The smaller size of ODE produces less drag and less engine heat transfer during hypersonic flight. It was proposed by Nicholls et al. in 1958 [1,2] and attracts much attention all over the world in recent years [3,4]. ODE is also called shock-induced combustion ramjet (Shcramjet) sometimes although they are not exactly the same [5]. Therefore, the research progress of ODE and Shcramjet are briefly reviewed together here.

In experiments, Rubins et al. were the first to conduct ODE experiments in a Mach 3 tunnel operating at temperatures in the range of 1944 K in 1959 [6-8]. They produced normal shock-induced combustion and oblique shock-induced combustion in experiments. They demonstrated that continuous supersonic combustion can be generated by shock heating an air-fuel mixture to the ignition static temperature in a supersonic flow. They also discussed the relation between shock-induced combustion and standing detonation waves.

Sterling et al. conducted a serious of ODE experiments in Caltech T-5 high enthalpy shock tunnel [9]. A hydrogen injection system located within the T-5 nozzle was used to provide hydrogen coflowing jet to premix with the hypersonic air upstream of a wedge model placed in the test section. The oblique detonation wave was initiated on the wedge. Morris et al. used OH planar laser-induced fluorescence (PLIF) and schlieren imaging to investigate shock-induced combustion phenomena on a 40° wedge in an expansion tube in Stanford University [10]. Stoichiometric $H_2/O_2$ gas mixtures with three different levels of nitrogen dilution were tested at two different test conditions in their experiments. The shock-induced combustion behind an attached shock at the tip of the wedge was yielded in the experiments.

Veraar et al. designed an axi-symmetric dual cone test object to demonstrate that it is possible to inject hydrogen into a high enthalpy supersonic air flow without causing premature ignition and subsequently induce combustion of this mixture by a strong oblique shock wave [11]. The experiments were executed in the TNO Free Jet Test Facility using the Mach 3.25 free jet nozzle. These experimental results were an important step in demonstrating the feasibility of a shock-induced combustion ramjet as a future hypersonic propulsion system.

Gong et al. studied the process of oblique detonation initiation through a wedge or cone in a supersonic wind tunnel [12]. The high-speed Schlieren images demonstrated the unstable feature of stationary oblique detonation wave. Rosato et al. designed a unique hypersonic high-enthalpy reaction facility that produces an oblique detonation wave stabilized in space. A standing oblique detonation wave was created in a hypersonic flow of hydrogen and air. This breakthrough allowed for a possible pathway to develop and integrate an ultra-high-speed ODE. Zhang et al. conducted the first free-jet experiments of a large-scale hydrogen-fueled oblique detonation engine model in JF-12 shock tunnel [14,15]. The results showed that the stabilized oblique detonation wave was established in the combustor successfully during the 50ms test time. Han et al. conducted the experiments with RP-3 kerosene fuel [16]. In order to overcome the difficulty of initiation of kerosene, a forced initiation technique was put forth and RP-3 kerosene oblique detonation was initiated successfully. Their experimental results further demonstrated the feasibility of ODE as a future hypersonic propulsion system.

In numerical simulations of ODE and Shcramjet, Sislian et al. were the first to conduct two-dimensional numerical simulations to study the propulsive performance of full-scale $H_2$/air ODE with Euler equations and nonequilibrium detailed chemical reaction kinetics [17-19]. The flight Mach number region was between Ma9 and Ma16. The incoming flow was assumed to be premixed stochiometric $H_2$/air mixture. The combustor entrance static temperature was set to be 900K. The specific impulse of ODE was obtained numerically. The results demonstrated that shock-induced combustion can be used as a means of hypersonic propulsion. The results also demonstrate that a minimum-entropy or Chapman-Jouguet condition exists for the oblique detonation wave generated by a wedge.

Alexander et al. performed three-dimensional numerical simulations of the full-scale $H_2$/air Shcramjet flow field at Ma11 and an altitude of 34.5 km with Favre-averaged Navier–Stokes equations and detailed chemical reaction kinetics [20-22]. The length and height of the Shcramjet were 4.170m and 0.465m, respectively. Cantilevered ramp injector arrays were used to mix the hydrogen with air in the isolator. The specific impulse of the engine is about 803s. Chan carried out numerical simulations and compared the propulsive performance of a Scramjet and Shcramjet at Ma11. The ODE obtained a specific impulse of 1290s at Ma11[23].

Ma et al. carried out the two-dimensional numerical simulations of Ma9 full-scale $H_2$/air Shcramjet at altitude of about 45km [24]. The length of the Shcramjet is 5.023m. Two struts were used to inject the hydrogen in the isolator. The mixing process was not good because of the two-dimensional simulation. Shock-induced combustion was obtained in the combustor. Zhang et al. performed numerical simulation of a Ma9 ODE [25]. Three struts were used to inject hydrogen at the leading edge of inlet. The mixing process was simulated by three-dimensional simulation and the combustion process of oblique detonation in the combustor was simulated by two-dimensional simulation. Oblique detonation wave induced by oblique shock wave and normal detonation wave induced by Mach stem were obtained in the combustor, respectively.

For the theoretical analysis of ODE propulsive performance, the analysis program of ODE is similar to that of Ramjet, Scramjet or RBCC, so the references of Ramjet, Scramjet and RBCC are also included in this section. The first Ramjet Performance Analysis Code (RJPA) was developed at Johns Hopkins University in the mid-1960's [26]. Morrison was the first to perform the theoretical analysis on the potential propulsive performance of ODE and obtained the thrust coefficients and specific impulses for hydrogen-air mixture [27, 28]. The conclusion was that ODE offers great potential as an air-breathing propulsor to extend the range of the ramjet flight Mach numbers from takeover speeds of Ma6 to Ma16. Ostrander et al. wrote two computer programs to model the ODE flow fields and calculate the propulsive performance [29]. Pratt et al. put forth the morphology of standing oblique detonation waves [30, 31].

The Simulated Combined-Cycle Rocket Engine Analysis Module (SCCREAM) was developed in Georgia Institute of Technology and was used to analysis the propulsive performance of Hyperion, an SSTO vision vehicle concept utilizing RBCC propulsion [32, 34]. Ashford and Emanuel evaluated the performance of ODE from Ma5 to Ma20 and compared it with Scramjet [35]. Valorani et al. presented a mathematical model to predict the overall performance of ODE [36]. Mogavero et al. described the initial steps towards the construction of an engineering tool, called the Hybrid Propulsion Parametric-Modular Model (HPPMM), which can be used to study the behavior of Scramjet through the use of simplified numerical models [37]. Zhang et al. theoretically analyzed the optimum performance of kerosene scramjet engine at high Mach numbers by using Brayton cycle [38]. Zhang et al. developed an analysis tool of the rocket-based combined cycle engine named Skye [39]. Yang et al. analyzed the overall propulsive performance of ODE [40]. In their analysis, four different combustion modes, over-driven ODW, Chapman-Jouguet ODW, over-driven normal detonation wave and oblique shock-induced constant-volume combustion were considered, respectively.

In summary, all of the theoretical analysis on the propulsive performance of ODE were done by writing a computer code on the basis of theorem of momentum. The aim of this research is to derive a mathematical equation of the propulsive performance of ODE, which can express the physical law and key parameters explicitly. A simplified physical model of ODE is put forth, the equation to calculate the thrust generated by the divergent nozzle is deduced, and the key parameters governing the propulsive performance are obtained. The specific impulse of ODE with hydrogen and kerosene fuel is calculated by this equation for the flight Mach number region between Ma8 and Ma15.

**2 The criterion on the Propulsive Balance of ODE**

**2.1 Physical model of ODE**

The physical model of ODE is given in Fig.1. The ODE consists of three parts, the inlet, the constant cross-section combustor and the divergent nozzle. The inlet compresses the incoming flow to an appropriate level by oblique shock waves and generates drag force. The oblique detonation wave is initiated in the constant cross-section combustor and the combustor does not produce any pressure force. The divergent nozzle generates thrust by expanding the supersonic high pressure and high temperature detonation product isentropically. We can see obviously that the gas dynamic characteristics of the flow field of these three parts are totally different. Therefore, we can model them separately by using different theory and then obtain the criterion on the propulsive balance of ODE.

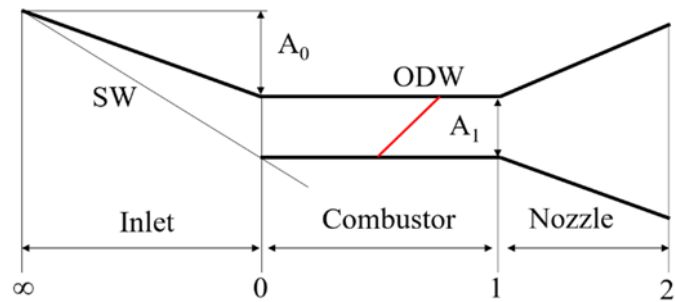

Fig.1 Physical model of ODE

In order to conduct theoretical analysis, some assumptions are made as follows.

(1) The ODE configuration is two-dimensional.

(2) The flow field is inviscid and the friction force is not considered.

(3) The inlet is of single-stage compression, the combustor is of constant cross-section, and the nozzle is a symmetric divergent nozzle.

(4) The lip of the cowl is aligned with the combustor entrance vertically.

(5) The engine operates at the design-point, that is, the inlet oblique shock wave strikes on the lip of cowl exactly.

(6) The flow at the combustor entrance is uniform without considering any reflection or compression of the flow in the isolator.

(7) The detonable mixture is premixed and assumed to be ideal gas.

(8) The C-J oblique detonation wave is initiated in the combustor ideally without any reflection and shock/shock interaction.

(9) The pressure drag produced by the wedge generating C-J oblique detonation wave is not considered.

## 2.2 The thrust of Laval nozzle

According to the classical gas dynamic theory, the variation of the static pressure and the area during the isentropic expansion in the Laval nozzle is expressed by Eq.(1) and Eq.(2).

$$\frac{p}{p_0} = \left(1 + \frac{\gamma-1}{2} Ma^2\right)^{-\frac{1}{\gamma-1}} \qquad (1)$$

$$\frac{A^*}{A} = \left(\frac{\gamma+1}{2}\right)^{\frac{\gamma+1}{2(\gamma-1)}} Ma \left(1 + \frac{\gamma-1}{2} Ma^2\right)^{-\frac{\gamma+1}{2(\gamma-1)}} \qquad (2)$$

where, $p$ is the static pressure, $p_0$ the total pressure, $Ma$ the exit Mach number, $A^*$ the throat area, $A$ the exit area, and $\gamma$ the specific heat ratio, respectively.

The thrust $F^*$ produced by the Laval nozzle is derived by integrating the pressure on the nozzle surface in Eq.(3).

$$\frac{F^*}{p^* A^*} = \left(\gamma Ma + \frac{1}{Ma}\right) \left[\frac{2}{\gamma+1}\left(1 + \frac{\gamma-1}{2} Ma^2\right)\right]^{-\frac{1}{2}} - \gamma - 1 \qquad (3)$$

It can be seen from Eq.(3) that there are three key parameters influencing the thrust $F^*$ generated by the Laval nozzle, the static pressure at throat $P^*$, the area of throat $A^*$, and the exit Mach number $Ma$. The exit Mach number also means the size of the nozzle.

The nondimensional thrust of Laval nozzle with different specific ratio calculated from Eq.(3) is shown in Fig.2. It can be seen that the thrust increases rapidly with exit Mach number at lower Mach number (Ma<5) and then becomes flat at higher Mach number. This characteristic can be demonstrated by Eq.(3) because the nondimensional thrust becomes Mach-independent at higher Mach number. This result means that there is an optimized nozzle size for air-breathing engine.

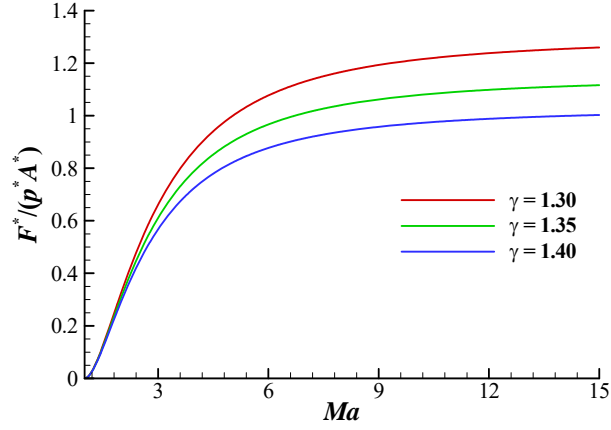

Fig.2 Nondimensional thrust of Laval nozzle with different specific heat ratio

### 2.3 The thrust of divergent nozzle

The nondimensional thrust of divergent nozzle (the subscript see Fig.1) given in Eq.(4) is easily obtained from Eq.(3). Note that the thrust in Eq.(4) is nondimensionalized by the theoretical throat parameters. The thrust can also be nondimensionalized by the divergent nozzle entrance parameters by the relationship of Eq.(5).

$$\frac{F}{p^*A^*} = \left(\gamma Ma_2 + \frac{1}{Ma_2}\right)\left[\frac{2}{\gamma+1}\left(1+\frac{\gamma-1}{2}Ma_2^2\right)\right]^{-\frac{1}{2}} - \left(\gamma Ma_1 + \frac{1}{Ma_1}\right)\left[\frac{2}{\gamma+1}\left(1+\frac{\gamma-1}{2}Ma_1^2\right)\right]^{-\frac{1}{2}} \quad (4)$$

$$p^*A^* = Ma_1\left[\frac{2}{\gamma+1}\left(1+\frac{\gamma-1}{2}Ma_1^2\right)\right]^{\frac{1}{2}} p_1 A_1 \quad (5)$$

where, $F$ is the thrust generated by the divergent nozzle, $Ma_1$ the entrance Mach number, and $Ma_2$ the exit Mach number, respectively.

It can be seen from Eq.(4) that there are four key parameters governing the divergent nozzle thrust, the entrance Mach number $Ma_1$, entrance static pressure $P_1$, entrance area $A_1$, and exit Mach number $Ma_2$. For an ODE, the nozzle entrance Mach number $Ma_1$ is calculated from the C-J oblique detonation theory. Therefore, we should know the exit Mach number $Ma_2$ for a given divergent nozzle. The exit Mach number $Ma_2$ is a function of the entrance Mach number $Ma_1$ and the expansion ratio $A_2/A_1$ of the divergent nozzle as shown in Eq.(6).

$$Ma_2 = f\left(Ma_1, \frac{A_2}{A_1}\right) \tag{6}$$

The nondimensional thrust of divergent nozzle under different expansion ratio cannot be expressed by a simple mathematical equation. It is numerically calculated for the specific ratio of $\gamma = 1.3$ and the result is plotted in Fig.3. We can obtain two important results from Fig.3. The first is that the nondimensional thrust of divergent nozzle increases with the nozzle entrance Mach number, which means that the supersonic combustion is good for the propulsive performance of hypersonic air-breathing engine. The second is that the nondimensional thrust does not increase obviously when the expansion ration $A_2/A_1$ is larger than 20. This is the optimized size of the divergent nozzle.

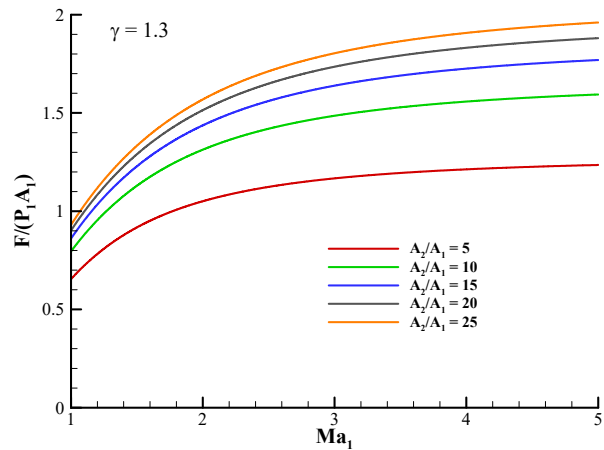

Fig.3 Nondimensional thrust of divergent nozzle under different expansion ratio

The simpler equations for engineering application from Eq.(7) through Eq.(10) are derived by fitting the numerical results for the range of nozzle entrance Mach number $Ma_1$ from 1 to 5. The term of $\ln(Ma_2/Ma_1)$ in Eq.(9) and Eq.(10) represents the size of the divergent nozzle. The thrust of a divergent nozzle with expansion ratio $A_2/A_1$ larger than 20 is given in Eq.(11). These equations demonstrate that the main method to increase the thrust is to increase the combustion pressure.

$$\frac{F^*}{p^* A^*} = 0.7446 \ln Ma - 0.132 \tag{7}$$

$$p^* A^* = 0.8835 Ma_1^{1.5339} p_1 A_1 \tag{8}$$

$$\frac{F}{p^* A^*} = 0.7446 \ln\left(\frac{Ma_2}{Ma_1}\right) \tag{9}$$

$$\frac{F}{p_1 A_1} = 0.6578 Ma_1^{1.5339} \ln\left(\frac{Ma_2}{Ma_1}\right) \tag{10}$$

$$\frac{F}{p_1 A_1} = 0.7862 \ln(Ma_1) + 1.0 \text{ for } A_2/A_1 \geq 20 \tag{11}$$

### 2.4 The criterion on the propulsive balance

The pressure drag generated by the single-stage compression inlet is given in Eq.(12).

$$D = p_0 A_0 \tag{12}$$

where, $D$ is the pressure drag of inlet, $P_0$ the static pressure at inlet exit, $A_0$ the projected area of inlet in the thrust direction (see Fig.1).

The criterion on the propulsive balance is obtained in Eq.(13) when the thrust generated by the divergent nozzle is equal to the pressure drag produced by the inlet.

$$(0.7862 \ln(Ma_1) + 1.0) p_1 A_1 = p_0 A_0 \tag{13}$$

The nondimensional net thrust $\bar{F}$ of ODE is calculated by Eq.(14), and $n$ is defined as the combustion pressure ratio in Eq.(15).

$$\bar{F} = \frac{F}{P_0 A_1} = (0.7862 \ln(Ma_1) + 1.0) n - A_0/A_1 \tag{14}$$

$$n = \frac{P_1}{P_0} \tag{15}$$

We can find from Eq.(14) that the combustion pressure ratio $n$ is a very important parameter for the net thrust of air-breathing engine. In order to increase the net thrust, on one hand, we should try to increase the combustion pressure ratio $n$ by using detonation combustion mode; on the other hand, try to decrease the compression ratio of inlet $A_0/A_1$ by using multiple-stage compression inlet. The combustion pressure ratio $n$ is a function of inlet exit parameters, fuel and equivalence ratio. The inlet exit static temperature $T_0$ influences the combustion pressure ratio $n$ very much, which is also a function of the inlet compression ratio $A_0/A_1$. Therefore, the two terms in the right hand of Eq.(14) are coupled together.

**3 Validation of the Equations**

The equation of Eq.(10) is validated by the numerical results of the two-dimensional numerical simulations of Ma9 full-scale $H_2$/air Shcramjet [24]. The total length of the full-scale engine is 5023mm. The single-compression inlet has a length of 2524mm and an deflection angle of 13°. The wedge in the combustor has an deflection angle of 20°. The thrust nozzle has a length of 1597mm and an expanded angle of 20°. The flight Mach number is M9 and the angle of attack is zero degree. The freestream conditions are at an altitude of 40km with a static pressure of 250Pa and a static temperature of 287K.

The temperature contours of the full-scale Shcramjet and the local enlarged combustor are shown in Fig.4 and Fig.5, respectively. The nozzle entrance and exit parameters used for theoretical analysis are obtained from the density-averaged parameters of numerical results. We can see from Fig.5 that the combustion flow field of the combustor is very complicated. So, three entrance sections are selected, x=0.88m, 0.89m, 0.90m, respectively. The exit section is set at x=1.5m (see Fig.4). The width of the engine is assumed to be 1m. The results are given in Table 1. The discrepancy of the nozzle thrust by theoretical analysis and CFD is 5.48%, -1.83%, -2.46%, respectively, which demonstrates that the theoretical thrust equation is correct.

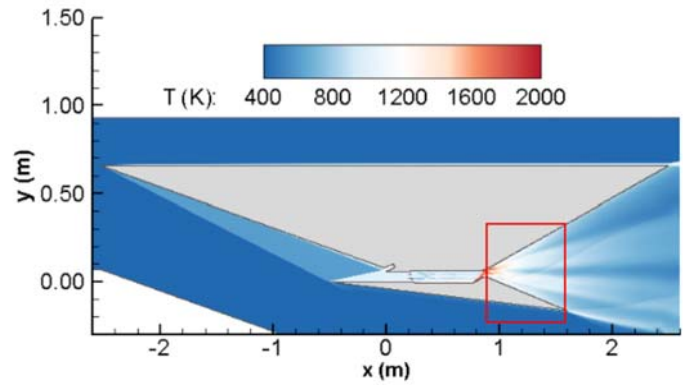

Fig.4 The pressure contours of the Shcramjet

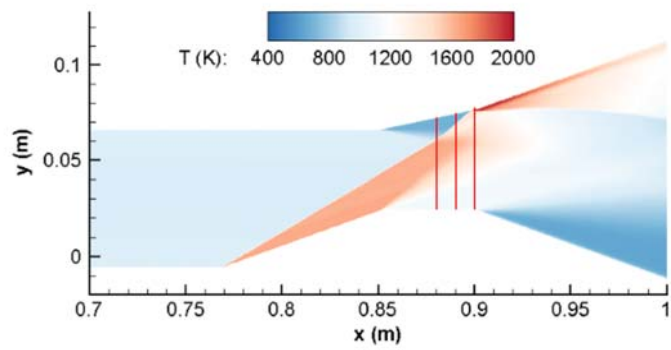

Fig.5 The temperature contours of the combustor

**Table 1 Companion of nozzle thrust by theoretical analysis and CFD**

| Parameters | x=0.88m | x=0.89m | x=0.9m |
| --- | --- | --- | --- |
| $P_1$ (Pa) | 41221 | 37533 | 34252 |
| $Ma_1$ | 2.35 | 2.51 | 2.57 |
| $A_1$(mm) | 47.8 | 49.8 | 54.3 |
| $P_2$ (Pa) | 1322 | 1322 | 1322 |
| $Ma_2$ | 4.35 | 4.35 | 4.35 |
| $A_2$(mm) | 493.9 | 493.9 | 493.9 |
| $F_{theory}$ (N) | 2963 | 2766 | 2734 |
| $F_{theory}/P_1A_1$ | 1.50 | 1.48 | 1.47 |
| $F_{CFD}$ (N) | 2802 | 2802 | 2802 |
| $F_{CFD}/P_1A_1$ | 1.422 | 1.498 | 1.506 |
| Discrepancy | 5.48% | -1.83% | -2.46% |

# 4 Propulsive Performance of ODE

## 4.1 The specific impulse of H₂/air mixture

The specific impulse of ODE with stoichiometric H₂/air mixture is calculated by the theoretical equations derived above. The calculation procedures are as follows.

(1) The gas is assumed to be premixed stoichiometric H₂/air mixture.

(2) The incoming flow parameters are $Ma_\infty$, $p_\infty$, $\rho_\infty$, $T_\infty = 250\text{K}$, $R_\infty = 397.7$, $MW = 20.9$, $\gamma = 1.3$. The values of static pressure and density are not necessary when calculating the specific impulse.

(3) The inlet exit static temperature is a very important parameter influencing the combustion pressure ratio. It is fixed to be $T_0 = 1000\text{K}$. The inlet exit Mach number $M_0$ and velocity $u_0$ can be calculated by classical oblique shock wave theory.

(4) The captured mass flow rate of ODE is $\dot{m}_{mix} = \rho_0 u_0 A_1$.

(5) The combustion pressure ratio $n$ and the Mach number of detonation product $M_1$ are calculated by C-J oblique detonation theory. The pressure of detonation product is $p_1 = n p_0 = n \rho_0 R_0 T_0$.

(6) The net thrust of ODE is

$$F = \left[ \left(0.7862 \ln(Ma_1) + 1.0 \right) n - A_0/A_1 \right] p_0 A_1 \\ = \left[ \left(0.7862 \ln(Ma_1) + 1.0 \right) n - A_0/A_1 \right] \rho_0 R_0 T_0 A_1.$$

(7) The mass flow rate of hydrogen is $\dot{m}_{H_2} = c_{H_2} \rho_0 u_0 A_1 = 0.029 \rho_0 u_0 A_1$.

(8) The specific impulse of ODE is

$$I_{sp} = \frac{\left[ \left(0.7862 \ln(Ma_1) + 1.0 \right) n - A_0/A_1 \right] \rho_0 R_0 T_0 A_1}{c_{H_2} \rho_0 u_0 A_1 g} \\ = \frac{\left[ \left(0.7862 \ln(Ma_1) + 1.0 \right) n - A_0/A_1 \right] R_0 T_0}{c_{H_2} u_0 g}.$$

The specific impulse of ODE in the flight Mach number region between Ma8 and Ma15 is theoretically calculated. The results are shown in Table 2 and Fig.6. The Mach number of C-J oblique detonation product is also given in Table 2. The pressure ratio of C-J oblique detonation is 4.46. The data from references are also plotted in Fig.6 for comparison. Note that the inlet exit static temperature is a very important parameter, and the inlet exit static temperature of references for comparison should be also about 1000K. It can be seen from Fig.6 that the ODE obtains positive net thrust. The specific impulse is about 1763s at Ma8 and about 1392s at Ma15, respectively. The results are in good agreement with references at lower Mach numbers, but larger at higher flight Mach numbers. In the two-dimensional numerical simulations of Sislian[19] and Dudebout[18], the pressure drag of the cowl was also considered; therefore, their specific impulse is smaller than that of this study.

**Table 2 The Specific Impulse under different flight Mach number**

| $Ma_\infty$ | 8 | 9 | 10 | 11 | 12 | 13 | 14 | 15 |
|---|---|---|---|---|---|---|---|---|
| $Ma_1$ | 1.77 | 2.14 | 2.57 | 2.91 | 3.28 | 3.6 | 3.99 | 4.31 |
| $A_0/A_1$ | 3.25 | 3.57 | 3.82 | 3.98 | 4.11 | 4.22 | 4.3 | 4.35 |
| $u_0$ (m/s) | 2551 | 2924 | 3362 | 3728 | 4130 | 4473 | 4905 | 5263 |
| $I_{sp}$ (s) | 1763 | 1703 | 1645 | 1587 | 1481 | 1481 | 1431 | 1392 |

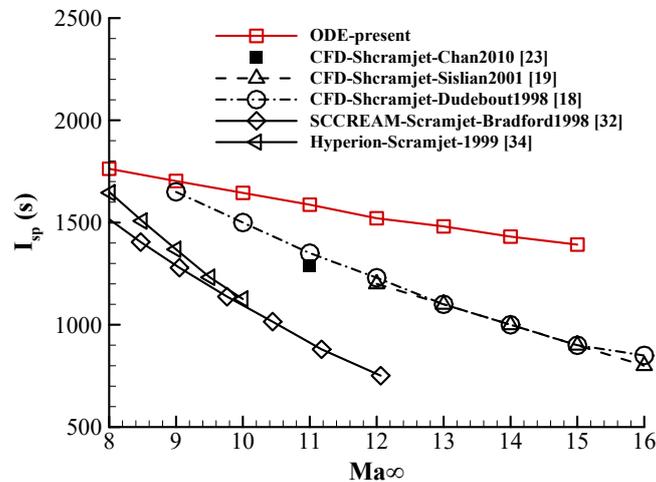

Fig.6 The specific impulse of ODE of stoichiometric $H_2$/air mixture

**4.2 The specific impulse of kerosene/air mixture**

The specific impulse of ODE with stoichiometric kerosene/air mixture is calculated by the same procedure and the results are plotted in Fig.7. The specific heat ratio is assumed to be $\gamma=1.4$. The inlet exit static temperature is also set to be 1000K. The ODE obtains positive net thrust from Ma8 to Ma15. The specific impulse is about 1000s at Ma8 and 800s at Ma15, respectively. The results are in good agreement with that of references, especially with the result of Ashford [35] at higher Mach numbers. The specific impulse of kerosene is about a half of that of hydrogen because the mass flow rate of kerosene is about two times of hydrogen. The combustion pressure ratio of kerosene/air is higher than hydrogen/air, while the gas constant is lower. Therefore, the influences of these two parameters on specific impulse is offset.

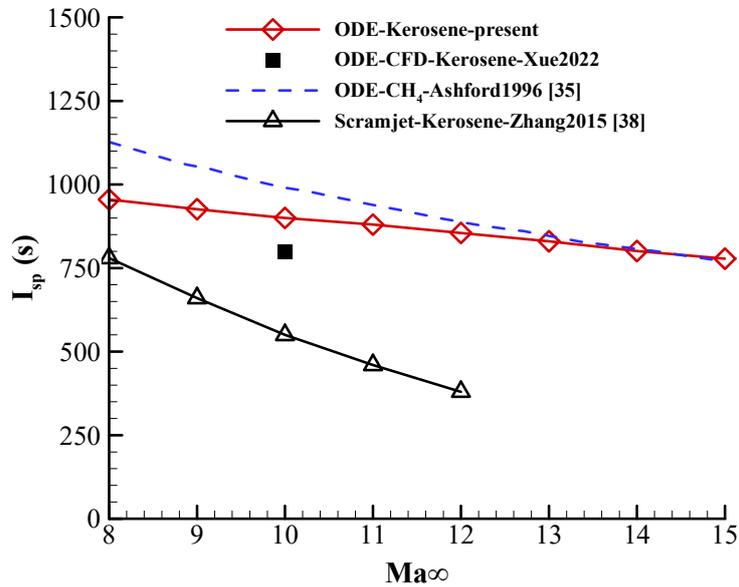

Fig.7 The specific impulse of ODE of stoichiometric kerosene/air mixture

**4.3 The influence of inlet exit temperature**

The inlet exit temperature $T_0$ influences the thrust of ODE very much. According to Eq.(14), the higher the inlet exit temperature $T_0$ is, the lower the combustion pressure ratio $n$ will be, the lower the detonation product Mach number $Ma_1$ will be, and the larger the inlet compression ratio $A_0/A_1$ will be. As a result, the lower the nondimensional net thrust will be.

The pressure ratio of C-J detonation of stoichiometric $H_2$/air mixture and stoichiometric hydrocarbon/air mixture under different static temperature is plotted in Fig.8 [24]. The theoretical results show that the static temperature at the entrance of combustor influences the combustion pressure ratio very much. For the $H_2$/air mixture, the pressure ratio is 15 at 300K and only 2.93 at 1500K, respectively. For the $C_8H_{18}$/air mixture, the pressure ratio is 17.48 at 300K and only 3.59 at 1500K, respectively. In order to increase the combustion pressure ratio, the static temperature at the entrance of combustor should be decreased. This result also indicates that the combustion pressure ratio of hydrocarbon/air is about 15% higher than that of hydrogen/air mixture, which means that hydrocarbon could produce larger thrust than hydrogen under the same conditions.

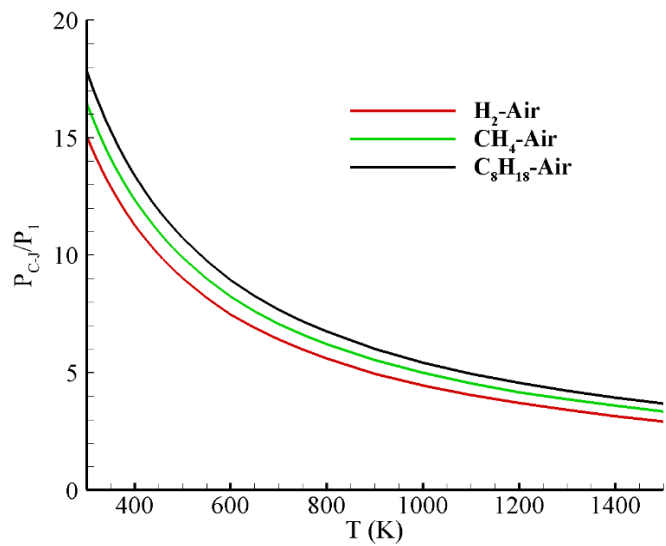

Fig.8 Pressure ratio of C-J detonation under different static temperature

The specific impulse of Ma12 ODE with stoichiometric $H_2$/air mixture under different inlet exit temperature from 600K to 1000K is plotted in Fig.9. We and find that the specific impulse decreases with the increase of temperature gradually. The present results are following the same trend as that of Sislian [19]. Note that the pressure drag of the cowl was also included in the two-dimensional numerical simulations of Sislian; therefore, his results are lower than the present study. The theoretical results are in good agreement with that of Ashford [35] and Yang [40].

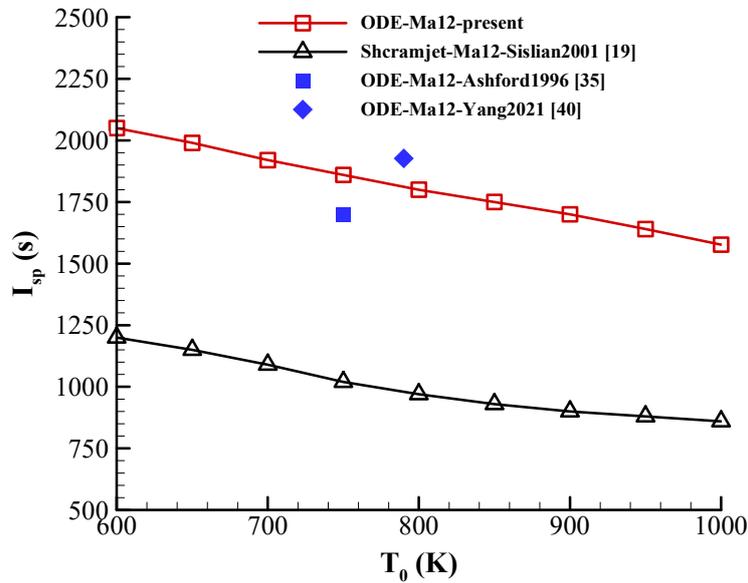

Fig.9 The specific impulse of Ma12 ODE under different inlet exit temperature

**5 Conclusions**

This theoretical analysis on the propulsive performance ODE comes to two conclusions.

(1) A physical model of ODE is put forth and mathematical equations of the criterion on the propulsive balance of ODE are deduced. The analysis demonstrates that the inlet exit temperature is a very important parameter influencing the propulsive performance of ODE. The lower the inlet exit temperature, the higher the specific impulse.

(2) The specific impulse of ODE in the flight Mach number region from Ma8 to Ma15 of stoichiometric $H_2$/air mixture and hydrocarbon/air mixture are calculated by using these equations. The results show that ODE can obtain positive net thrust in hypersonic flight region, which demonstrate the superiority of ODE for hypersonic flight.

**Acknowledgements**

This study is founded by the National Natural Science Foundation of China (No.11672312).

**References**

[1] J.A. Nicholls, E.K. Dabora, R.L. Gealer, Studies in connection with stabilized gaseous detonation waves. Proc Combust Inst. 7 (1) (1958) 766-772.

[2] R. Dunlap, R.L. Brehm and J.A. Nicholls, A preliminary study of the application of steady-state detonative combustion to a reaction engine. Journal of Jet Propulsion. 28 (7) (1958) 451-456.

[3] K. Kailasanath, Review of propulsion applications of detonation waves. AIAA Journal. 38 (2000) 1698-1708.

[4] P. Wolański, Detonative propulsion. Proceedings of the Combustion Institute. 34 (2013) 125-158.

[5] W. Huang, H. Qin, S. Bin, et al., Research status of key techniques for shock-induced combustion ramjet (shcramjet) engine. Science China Technological Sciences. 53 (1) (2010) 220-226.

[6] P.M. Rubins, R.P. Rhodes, Shock-induced combustion with oblique shocks, comparison of experiment and kinetic calculations. AIAA Journal. 1(12) (1963) 2778-2784.

[7] P.M. Rubins, and T.H.M. Cunningham, Shock-induced combustion in a constant area duct. Journal of Spacecraft and Rockets. 2 (2) (1965) 199-205.

[8] P.M. Rubins, R.C. Bauer, Review of shock-induced supersonic combustion research and hypersonic applications. Journal of Propulsion and Power. 10(5) (1994) 593-601.

[9] J.D. Sterling, E.B. Cummings, K. Ghorbanian, et al., Oblique detonation wave studies in the Caltech T-5 shock tunnel facility. AIAA-98-1561, 1998.

[10] C.I. Morris, M.R. Kamel, R.K. Hanson, Shock-induced combustion in high-speed wedge flows. Symposium on Combustion. 27(2) (1998) 2157-2164.

[11] R.G. Veraar, A.E.H.J. Mayer, J. Verreault, et al., Proof-of-principle experiment of a shock-induced combustion ramjet. AIAA-2009-7432, 2009.

[12] J.S. Gong, Y.N. Zhang, H. Pan, et al., Experimental investigation on initiation of oblique detonation waves. AIAA-2017-2350, 2017.

[13] D.A. Rosato, M. Thorntona, J. Sosa, et al., Stabilized detonation for hypersonic propulsion. Proceedings of the National Academy of Sciences. 118 (2) (2021) e2102244118.

[14] Z.J. Zhang, X. Han, K.F. Ma, et al., Experimental research on combustion mechanism of oblique detonation engines. Journal of Propulsion Technology. 42(4) (2021) 786-794.

[15] Z.J. Zhang, C.Y. Wen, C.K. Yuan, et al., An experimental study of formation of stabilized oblique detonation waves in a combustor. Combustion and Flame. 237 (2022) 111868.

[16] X. Han, W.S. Zhang, Z.J. Zhang, et al., Experimental study on RP3 aviation kerosene oblique detonation engine. Journal of Experiments in Fluid Mechanics. https://kns.cnki.net/kcms/detail/11.5266.V.20221110.1516.002.html

[17] R. Dudebout, J.P. Sislian, and R. Oppitz, Numerical simulation of hypersonic shock-induced combustion ramjets. Journal of Propulsion and Power, 14(6) (1998) 869-879.

[18] J.P. Sislian, R. Dudebout, J. Schumacher, et al., Incomplete mixing and off-design effects on shock-induced combustion ramjet performance. Journal of Propulsion and Power. 16(1) (2000) 41-48.

[19] J.P. Sislian, H. Schirmer, R. Dudebout, et al., Propulsive performance of hypersonic oblique detonation wave and shock-induced combustion ramjets. Journal of Propulsion and Power. 17(3) (2001) 599-604.

[20] D.C. Alexander, J.P. Sislian, B. Parent, Hypervelocity fuel/air mixing in mixed-compression inlets of shcramjets. AIAA Journal. 44(10) (2006) 2145-2155.

[21] J.P. Sislian, R.P. Martens, T.E. Schwartzentruber, et al., Numerical simulation of a real shcramjet flowfield. Journal of Propulsion and Power. 22(5) (2015) 1039-1048.

[22] D.C. Alexander, J.P. Sislian, Computational study of the propulsive characteristics of a shcramjet engine. Journal of Propulsion and Power. 24 (1) (2015) 34-44.

[23] J. Chan, Numerically simulated comparative performance of a scramjet and shcramjet at Mach 11. Thesis of Degree of Masters, University of Toronto, 2010.

[24] K.F. Ma, Z.J. Zhang, Y.F. Liu, et al., Aerodynamic principles of shock-induced combustion ramjet engines. Aerospace Science and Technology. 103 (2020) 105901.

[25] Z.J. Zhang, K.F. Ma, W.S. Zhang, et al., Numerical investigation of a Mach 9 oblique detonation engine with fuel pre-injection. Aerospace Science and Technology. 105 (2020) 106054.

[26] P. Pandolfini, Instructions for using Ramjet Performance Analysis (RJPA) IBM-PC Version 1.24, JHU/APL AL-92-P175. June 1992.

[27] R.B. Morrison, Evaluation of the oblique detonation wave ramjet. NASA TR-145358, 1978.

[28] R.B. Morrison, Oblique detonation wave ramjet. NASA CR-159192, 1980.

[29] M.J. Ostrander, J.C. Hyde, M.F. Young, et al., Standing oblique detonation wave engine performance. AIAA-87-2002, 1987.

[30] D.T. Pratt, J.W. Humphrey, D.E. Glenn, Morphology of a standing oblique detonation wave. AIAA-87-1785, 1987.

[31] D.T. Pratt, J.W. Humphrey, D.E. Glenn, Morphology of standing oblique detonation waves. Journal of Propulsion and Power. 7(5) (1991) 837-845.

[32] J.E. Bradford, J.R. Olds, Improvements and enhancements to SCCREAM, a conceptual RBCC engine analysis tool. AIAA-98-3775, 1998.

[33] J.E. Bradford, J.R. Olds, SCCREAM v.5: A web-based airbreathing propulsion analysis tool. AIAA-99-2104, 1999.

[34] J. Olds, J. Bradford, A. Charania, et al., Hyperion: an SSTO vision vehicle concept utilizing rocket-based combined cycle propulsion. AIAA-99-4944, 1999.

[35] S.A. Ashford, G. Emanuel, Oblique detonation wave engine performance prediction. Journal of Propulsion and Power. 12(2) (1996) 322-327.

[36] M. Valorani, M.D. Giacinto, C. Buongiorno, Performance prediction for oblique detonation wave engines (odwe). Acta Astronautica. 48(4) (2001) 211-228.

[37] A. Mogavero, I. Taylor, R.E. Brown, Hybrid propulsion parametric and modular model: a novel engine analysis tool conceived for design optimization. AIAA International Space Planes and Hypersonic Systems and Technologies Conference, 2015.

[38] D. Zhang, S. Yang, S. Zhang, et al., Thermodynamic analysis on optimum performance of scramjet engine at high Mach numbers. Energy. 90 (2015) 1046-1054.

[39] T. Zhang, Z. Wang, W. Huang, et al., An analysis tool of the rocket-based combined cycle engine and its application in the two-stage-to-orbit mission. Energy. 193 (2020) 116709.

[40] P.F. Yang, Z.J. Zhang, R.X. Yang, et al., Theorical study on propulsive performance of oblique detonation engine. Chinese Journal of Theoretical and Applied Mechanics. 53(10) (2021) 2853-2864.